\documentclass[review]{elsarticle}

\usepackage{lineno,hyperref}
\usepackage{graphicx}
\usepackage{amsmath}
\usepackage{amssymb}
\usepackage{tipa}
\usepackage{enumerate}
\usepackage{booktabs}
\usepackage{array}
\usepackage{colortbl}
\usepackage{float}

\modulolinenumbers[5]

\journal{Journal of \LaTeX\ Templates}

\newlength{\onesixth}
\newcolumntype{C}{>{\begin{minipage}{2\onesixth}\begin{center}}{c}<{\end{center}\end{minipage}}}
\newcolumntype{D}{>{\begin{minipage}{3\onesixth}\begin{center}}{c}<{\end{center}\end{minipage}}}
\newcolumntype{E}{@{}l@{}}

\begin{document}

\begin{frontmatter}

\title{ Experimental analyses on 2-hop-based and 3-hop-based link prediction algorithms}
\author[inst1,inst2]{Tao Zhou}
\author[inst1,inst2]{Yan-Li Lee\corref{cor1}}
\cortext[cor1]{Corresponding author.}
\ead{yanlicomplex@gmail.com}
\author[inst3]{Guannan Wang}

\address[inst1]{CompleX Lab, University of Electronic Science and Technology of China, Chengdu 611731, P.R. China}
\address[inst2]{Big Data Research Center, University of Electronic Science and Technology of China, Chengdu 611731, P.R. China}
\address[inst3]{Ant Financial Services Group, Hangzhou, 310099, People's Republic of China}

\begin{abstract}
Link prediction is a significant and challenging task in network science. The majority of known methods are similarity-based, which assign similarity indices for node pairs and assume that two nodes of larger similarity have higher probability to be connected by a link. Due to their simplicity, interpretability and high efficiency, similarity-based methods, in particular those based only on local information, have already found successful applications on disparate fields. In this research domain, an intuitive consensus is that two nodes sharing common neighbors are very likely to have a link, while some recent evidences argue that the number of 3-hop paths more accurately predicts missing links than the number of common neighbors. In this paper, we implement extensive experimental comparisons between 2-hop-based and 3-hop-based similarity indices on 128 real networks. Our results indicate that the 3-hop-based indices perform slightly better with a winning rate about 55.88\%, but which index is the best one still depends on the target network. Overall speaking, the class of Cannistraci-Hebb indices performs the best among all considered candidates.
\end{abstract}

\begin{keyword}
\texttt Complex Networks\sep Link Prediction\sep Similarity Index
\end{keyword}

\end{frontmatter}

\section{Introduction}

Link prediction is an elemental challenge in network science, which aims at estimating the existence likelihood of any nonobserved link, on the basis of observed links \cite{Getoor2005,Lu2011,Brugere2018,Squartini2018}. Theoretically speaking, link prediction can be treated as a testing aid for mechanism models, since a model that can well explain the network formation and evolution could be in principle transferred to an accurate link prediction algorithm \cite{WangEPL2012,WangEPL2014,ZhangSR2015, ZhangManag2017}. Practically speaking, prediction results can be used as an experimental guidance, by which we can focus on those biological interactions (e.g., regulatory interactions \cite{BarzelNB2013}, drug-target interactions \cite{DingBB2013}, protein-protein interactions \cite{ZhaoSR2015}) most likely to exist instead of blindly check all possible interactions, and thus the experimental costs can be largely reduced. Besides missing link problems, link prediction algorithms can also forecast links that may appear in the future of evolving networks, with significant commercial values in friend recommendations of online social networks \cite{AielloATW2012} and product recommendations in e-commercial web sites \cite{LuPR2012}.

Many algorithms have been proposed, including similarity-based algorithms \cite{LibenJAIST2007,ZhouEPJB2009}, probabilistic models
\cite{NevilleJMLR2007,YuNIPS2007,WangICDM2007}, maximum likelihood methods \cite{ClausetNature2008,GuimeraPNAS2009,PanSR2016}, and some other representatives \cite{LuPNAS2015,RathaEPL2017,BensonPNAS2018}. The probabilistic models and maximum likelihood methods are usually more accurate than similarity-based algorithms, at the same time, they suffer some intrinsic disadvantages. The probabilistic models often require information about node attributes in addition to the network structure, which highly limits their applications. Moreover, the number of parameters to be fixed are too many so that we cannot get insights about the network organization even if we have built a very accurate model. The maximum likelihood methods are highly time consuming, usually only capable to handle networks with a few thousands of nodes, while many real networks scale from millions to billions of nodes. Therefore, overall speaking, the similarity-based algorithms, in particular the ones based solely on local topological information, have found widest applications.

Generally speaking, a similarity-based algorithm will assign a similarity score to each pair of nodes, and assume that two nodes having a higher similarity score are of a larger likelihood to have a link. Therefore, all nonobserved links are ranked by their corresponding similarity scores, and the links with the highest scores are the predicted ones. Given a node pair $(i,j)$, many known local similarity indices only make use of the information contained in the 2-hop paths connecting $i$ and $j$, such as the common neighbor (CN) index \cite{LibenJAIST2007}, the resource allocation (RA) index \cite{ZhouEPJB2009}, the Adamic-Adar (AA) index \cite{AdamicSocNetw2003}, and the Cannistraci resource allocation (CRA) index \cite{CannistraciSR2013}. Besides different mathematical details, all 2-hop-based algorithms tend to assign a larger similarity score $S_{ij}$ if $i$ and $j$ have more 2-hop paths (i.e., more common neighbors). This is in accordance with an important network organization mechanism named as homophily \cite{McPhersonSocio2001,PanSR2016}, that is to say, two nodes having similar attributes are likely to connect to each other. In a network where only topological information is observed, the homophily mechanism can be interpreted as the fact that two nodes sharing one or more common neighbors are likely to become direct neighbors in the future. Such mechanism has been observed in the evolving processes of many real networks \cite{McPhersonSocio2001,KossinetsScience2006,RomeroAAAI2010,YinSIGIR2011,MaSR2016}, for example, more than 90\% of new links in Twitter and Weibo are between nodes that are already connected by at least one 2-hop path \cite{RomeroAAAI2010,YinSIGIR2011}. In a word, 2-hop paths are well accepted as strong evidence indicating the existence of missing link or future link between the corresponding two ends. The roles of longer paths are intuitively considered to be less significant since interacting strengths will decay along the paths \cite{KatzPsy1953,ChristakisNEJM2007}. Although a certain local similarity index (named as local path index) \cite{ZhouEPJB2009,LuPRE2009} has considered both contributions from 2-hop paths and 3-hop paths, the authors argued that the 2-hop paths play the leading role and the number of 3-hop paths plays a part only if the number of 2-hop paths is not sufficiently distinguishable.

Very recently, some scientists have argued that 3-hop-based similarity indices perform better than 2-hop-based indices \cite{RathaPA2019,KovacsNC2019,MuscoloniBioRxiv}. For example, Kov\'acs \emph{et al.} \cite{KovacsNC2019} proposed a degree-normalized index based on 3-hop paths and showed its remarkable advantage compared with 2-hop-based indices in predicting protein-protein interactions. Pech \emph{et al.} \cite{RathaPA2019} provided a theory showing that the number of 3-hop paths is a degenerated index of a more complicated index resulted from a linear optimization. Based on experimental analysis on eight networks, they argued that even the direct count of 3-hop paths performs better than the common neighbor index (i.e., the direct count of 2-hop paths), which is to some extent counterintuitive.

To clarify this issue, this paper implements experimental analyses on 128 real networks from 16 disparate fields. Extensive comparisons between 2-hop-based and 3-hop-based similarity indices indicate that the 3-hop-based indices perform slightly better with a winning rate about 55.88\%. However, given a specific target network, which index is the best choice still largely depends on the network structure. Overall speaking, the class of Cannistraci-Hebb (CH) indices \cite{MuscoloniBioRxiv} performs the best among all considered candidates, and the class of resource allocation (RA) indices \cite{ZhouEPJB2009} is the runner-up.

\section{Methods}

Denote $A$ as the adjacency network of a simple network $G$, where the element $a_{ij}=1$ if $i$ and $j$ are neighboring, and $a_{ij}=0$ otherwise. The degree of node $i$ is denoted by $k_i$ and the set of neighbors of node $i$ is denoted by $\Gamma_i$. Four representative 2-hop-based indices, CN \cite{LibenJAIST2007}, RA \cite{ZhouEPJB2009,Ou2007}, AA \cite{AdamicSocNetw2003}, and CH2 \cite{CannistraciSR2013,MuscoloniBioRxiv}, as well as their 3-hop-based counterparts are considered in this paper. To be clear, the suffixes L2 and L3 stand for 2-hop and 3-hop, for example, RA index and its 3-hop-based counterpart will be renamed as RA-L2 and RA-L3 indices.

CN-L2 index \cite{LibenJAIST2007} is a structural equivalence index. Two nodes are considered to be structural equivalence if they share many common neighbors. Accordingly, the similarity score between nodes $i$ and $j$ is
\begin{equation}
s_{ij}^{\rm{\text{CN-L2}}} = |\Gamma_i \cap \Gamma_j|.
\end{equation}

\begin{figure}[H]
\setlength{\abovecaptionskip}{0.cm}
\setlength{\belowcaptionskip}{-0.cm}
\centering
	\includegraphics[width=1\textwidth]{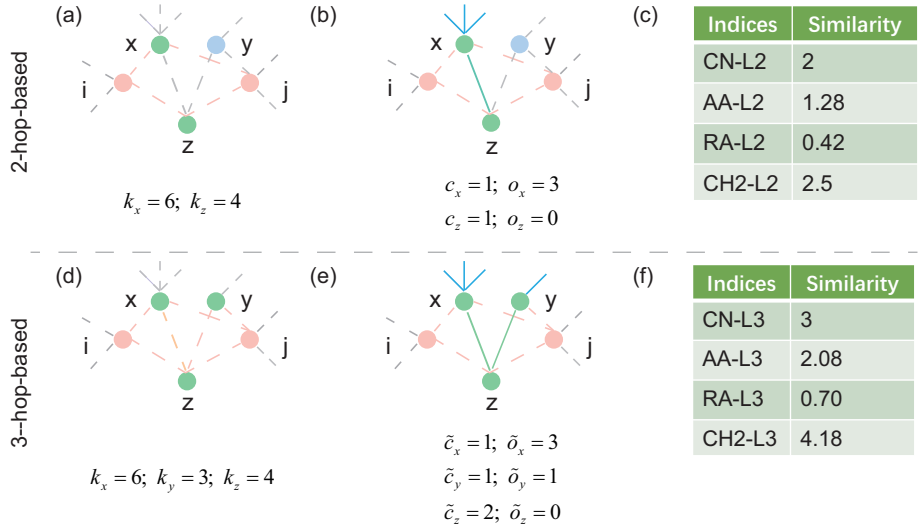}
    \caption{The illustration of eight similarity indices. Red nodes denote the target node pair, and green nodes denote intermediate nodes on 2-hop paths or 3-hop paths. Figure 1(a) is the illustration for indices CN-L2, AA-L2, RA-L2, where red links form 2-hop paths between nodes $i$ and $j$. Figure 1(b) is the illustration for CH2-L2, where green links and blue links denote the links counted by $c.$ and $o.$ in Eq. (\ref{CH2-L2}), respectively. Figure 1(d) is the illustration for indices CN-L3, AA-L3, RA-L3, where red links form 3-hop paths between nodes $i$ and $j$. Figure 1(e) is the illustration for CH2-L3, where green links and blue links denote links counted by $\tilde{c}.$ and $\tilde{o}.$ in Eq. (\ref{CH2-L3}), respectively. The corresponding similarity scores are shown in Figure 1(c) and Figure 1(f).}
\label{fig_IIlustration}
\end{figure}

AA-L2 index \cite{AdamicSocNetw2003} weakens the contribution of large-degree common neighbors, because to be neighboring to a popular node is generally less meaningful. The corresponding similarity score between nodes $i$ and $j$ is defined as
\begin{equation}
s_{ij}^{\rm{\text{AA-L2}}} = \sum_{x \in \Gamma_i \cap \Gamma_j}\frac{1}{\log (k_x)}.
\end{equation}

RA-L2 index \cite{ZhouEPJB2009,Ou2007} treats the similarity between nodes $i$ and $j$ as the resource transmitted from $i$ to $j$. Each neighbor of $i$ occupies one unit of resources, and allocates the resource equally to their neighbors. The resource received by node $j$ from $i$ is
\begin{equation}
s_{ij}^{\rm{\text{RA-L2}}} = \sum_{x \in \Gamma_i \cap \Gamma_j}\frac{1}{k_x}.
\end{equation}

CH2-L2 index \cite{MuscoloniBioRxiv} rewards internal links among common neighbors while penalizes links connecting common neighbors and outside. The similarity between nodes $i$ and $j$ is
\begin{equation}
\label{CH2-L2}
s_{ij}^{\rm{\text{CH2-L2}}}= \sum_{x \in \Gamma_i \cap \Gamma_j}\frac{1+c_x}{1+o_x},
\end{equation}
where $c_x$ is the number of $x$'s neighbors that are also in $\Gamma_i \cap \Gamma_j$, and $o_x$ is the number of $x$'s neighbors not in $\Gamma_i \cap \Gamma_j$, and not $i$ or $j$.

Correspondingly, CN-L3 index, AA-L3 index and RA-L3 \cite{KovacsNC2019} index are defined as follows,
\begin{equation}
s_{ij}^{\rm{\text{CN-L3}}} = \sum_{x \in \Gamma_i, y \in \Gamma_j}a_{xy},
\end{equation}
\begin{equation}
s_{ij}^{\rm{\text{AA-L3}}} = \sum_{x \in \Gamma_i, y \in \Gamma_j}\frac{a_{xy}}{\sqrt{\log(k_x)\log(k_y)}},
\end{equation}
\begin{equation}
s_{ij}^{\rm{\text{RA-L3}}} = \sum_{x \in \Gamma_i, y \in \Gamma_j}\frac{a_{xy}}{\sqrt{k_x k_y}}.
\end{equation}

CH2-L3 index \cite{MuscoloniBioRxiv} is defined as
\begin{equation}
\label{CH2-L3}
s_{ij}^{\rm{\text{CH2-L3}}} = \sum_{x \in \Gamma_i, y \in \Gamma_j}\frac{a_{xy}\sqrt{(1+\tilde{c}_x)(1+\tilde{c}_y)}}{\sqrt{(1+\tilde{o}_x)(1+\tilde{o}_y)}}.
\end{equation}
Similarly, $\tilde{c}_x$ is the number of links between $x$ and nodes in the set of intermediate nodes on all 3-hop paths connecting nodes $i$ and $j$, $\tilde{o}_x$ is the number of links between $x$ and nodes that are not $i$, $j$ or the intermediate nodes on any 3-hop paths connecting $i$ and $j$.

The illustration of the above eight indices are shown in figure \ref{fig_IIlustration}.

\section{Results}

\begin{figure}[H]
\setlength{\abovecaptionskip}{0.cm}
\setlength{\belowcaptionskip}{-0.cm}
\centering
	\includegraphics[width=1\textwidth]{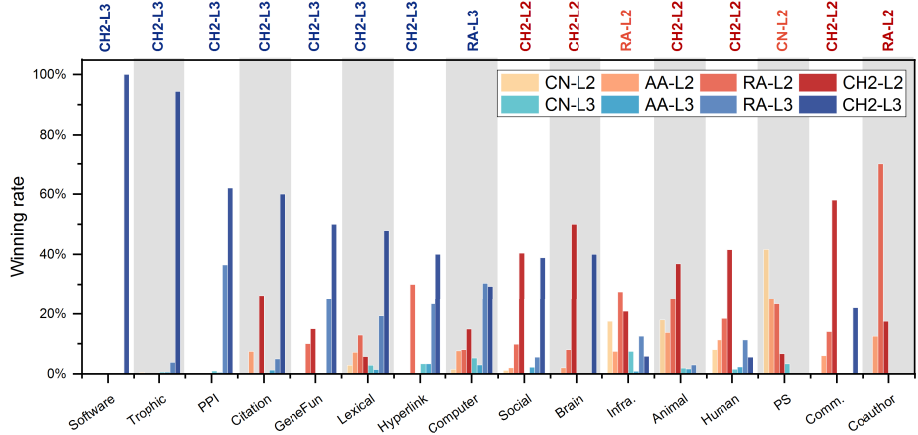}
    \caption{The winning rate of each similarity index for each network category. The 16 categories are shown in alternating white and grey for better discernibility. The 16 categories are: (1) Software \cite{Konect}---the networks of software components; (2) Trophic \cite{Gene2015,Pajek2006}---the predation networks of biological species; (3) PPI \cite{KovacsNC2019,Konect}---the protein-protein interaction networks; (4) Lexical \cite{Konect,MiloScience2004}---the networks of words from natural languages; (5) GeneFun \cite{Gene2015}---the networks of co-functional associations of genes; (6) Computer \cite{Konect,snap2014,circuits1989}---the networks of computer-related components; (7) Citation \cite{Konect,Pajek2006,TangKDD2009}---the citation networks; (8) Animal \cite{Konect}---the contact networks of animals; (9) Brain \cite{Gene2015,Wang2019}---the brain connection networks of cortical areas; (10) Infra. \cite{Konect}---the networks of physical infrastructures; (11) Hyperlink \cite{Konect}---the networks of web pages; (12) Social \cite{Konect,snap2014,MiloScience2004,Hu2019}---the online social networks; (13) Comm. \cite{Konect,Boguna2004}---the communication networks; (14) Coauthor \cite{snap2014}---the coauthorship networks; (15) Human \cite{Konect,Gene2015,GirvanPNAS2002}---the contact networks of human beings; (16) PS \cite{MiloScience2004}---the protein structure networks of secondary-structure elements. The best-performed index in each category is labelled above the corresponding window.}
\label{fig_field}
\end{figure}

Given a simple network $G(V, E)$, where $V$ is the set of nodes and $E$ is the set of links. To test the algorithm's accuracy, the set of links $E$ is randomly divided into two parts: (i) the training set $E^T$, which is the known information, and (ii) the probe set $E^P$, which is treated as the set of missing links. No information in $E^P$ is allowed to be used for the calculation of similarity matrix $S$. We adopt a standard metric, precision \cite{HerlockerATIS2004}, to quantify the algorithm's accuracy, which is defined as the ratio of the number of relevant elements in $S$ to the number of selected elements. In other words, if we select top-$L$ links as predicted links (i.e., the $L$ links with highest similarity scores in the set $E \verb|\| E^T$), among which $L_r$ links are in the probe set $E^P$, then the corresponding precision is $L_r/L$. In the experiments, for each network, the ratio of probe links to total links, say $q = |E^P|/|E|$, varies from 0.02 to 0.2, as $q = \{0.02, 0.04, \cdots ,0.2\}$. Correspondingly, we set the number of selected links $L = |E^P|$. 128 real networks from 16 disparate fields are used in the experimental comparisons, and thus 1280 comparisons among the eight similarity indices are recorded. The descriptions and topological statistics of these networks are shown in Supplementary Material Section S1.

\begin{figure}[t]
\setlength{\abovecaptionskip}{0.cm}
\setlength{\belowcaptionskip}{-0.cm}
\centering
	\includegraphics[width=1\textwidth]{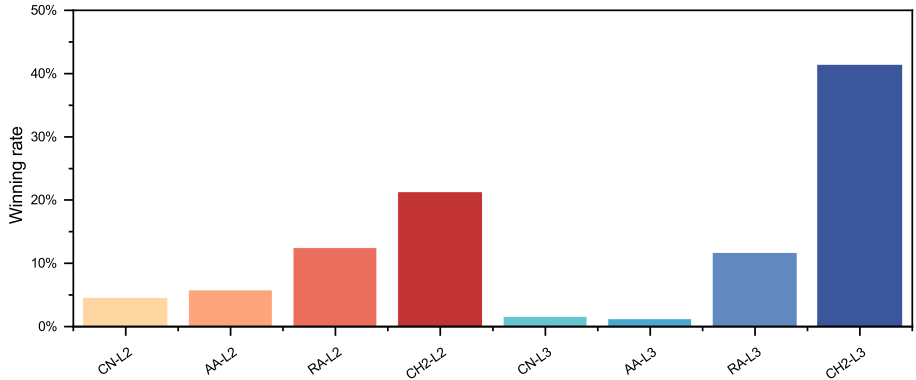}
    \caption{The overall winning rates of eight similarity indices in the 1280 comparisons.}
\label{fig_alg}
\end{figure}

Figure \ref{fig_field} shows the winning rate of each index for each network category. In a comparison, the best-performed index will get score 1 and all others get 0. If two indices are equally best, they both get score 0.5. The case with multiple winners is similar. The winning rate of an index is its total score divided by the number of comparisons. For example, in the category Hyperlink, there are three networks AB, BG and FD, and thus 30 comparisons. The index CH2-L3 wins 12 comparisons and thus has a high winning rate 40\%. As shown in figure \ref{fig_field}, in some categories the 3-hop-based indices are remarkably more accurate, while in some categories the 2-hop-based indices are much better. Yet for the category Social, the 2-hop-based and 3-hop-based indices exhibit nearly the same performance. Detailed results for each of the 128 networks are presented in Supplementary Material Section S2.

\begin{figure}[t]
\setlength{\abovecaptionskip}{0.cm}
\setlength{\belowcaptionskip}{-0.cm}
\centering
	\includegraphics[width=1\textwidth]{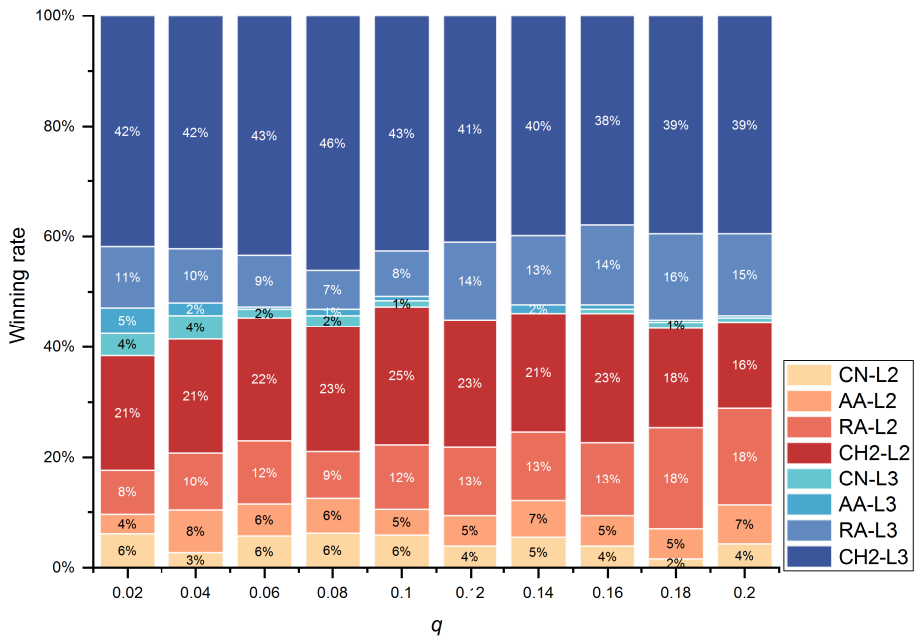}
    \caption{The winning rates of eight similarity indices for different $q$.}
\label{fig_q}
\end{figure}

As shown in figure \ref{fig_alg}, the overall winning rate for 3-hop-based indices in the 1280 comparisons is 55.88\%, and that for 2-hop-based indices is 44.12\%, namely the 3-hop-based indices perform slightly better as a whole. Together with the results reported in figure \ref{fig_field}, we can conclude that there is no easy way to anticipate which category is better or which index is the best. Indeed, which index is the best choice largely depends on the specific structure of the target network. An unexpected gain from these experiments is that the class of CH2 indices \cite{{MuscoloniBioRxiv}} perform the best among all 4 classes with a dominant winning rate 62.65\%, and the class of RA indices \cite{ZhouEPJB2009} is the runner-up with a winning rate 24.15\%, while the other two classes perform poorly, with both winning rates being about 7\%. Figure \ref{fig_q} validates the robustness of the above experimental results by varying the ratio $q$. Obviously, the main findings keep unchanged for different $q$.

\section{Discussion}

Based on extensive experiments, this paper provides a direct response to a recent debate about the roles of 2-hop paths and 3-hop paths on link prediction \cite{RathaPA2019,KovacsNC2019,MuscoloniBioRxiv}. The answer is not a simple winner, but a fact that 2-hop-based and 3-hop-based indices are competitive to each other. Indeed, the 3-hop-based indices perform slightly better as a whole, while which index is the best choice still largely depends on the specific structural features of the target network.

Such experimental observations immediately raise two new questions. Firstly, can we foreknow which index or which category of indices is better for a given network by measuring some structural features (of course, the computational complexity should be lower than direct comparisons of those indices)? Secondly, how to properly make use of information contained in both 2-hop and 3-hop paths to improve the algorithm's accuracy (at least a more subtle and effective way than the simply linear combination of CN-L2 and CN-L3 indices \cite{ZhouEPJB2009,LuPRE2009})? We leave these two open questions for future studies.

The longer paths are also relevant. For example, as suggested by Pech \emph{et al.} \cite{RathaPA2019}, to eliminate the redundancy in 3-hop-based indices by considering the 5-hop-based paths can further improve the algorithm's accuracy (this idea is very similar to a previous work \cite{ZhouNJP2009}). However, to account for longer paths is highly time-consuming while the improvement may be marginal. Intuitively, we do not think to consider longer paths is cost-efficient, however, intuition usually leads to mistakes, and thus whether our judgment is reasonable still needs further investigations.

An unexpected gain in this work is that the Cannistraci-Hebb indices \cite{MuscoloniBioRxiv} perform remarkably better than other indices. This is not a coincidence, but shows us an insight that the local connecting patterns in the neighborhood provide important information about the potential relationship between two nodes. In despite of the excellent performance of CH2-L3 index, it is just a naive extension of the CH2-L2 index. Once we known the value of the class of Cannistraci-Hebb indices, we are inspired to design more elegant and effective indices on the basis of local connecting patterns.

\section*{Acknowledgments}
The authors acknowledge the valuable discussion with Dr. Carlo Vittorio Cannistraci, and Dr. Qian-Ming Zhang for providing us some of the datasets. This work was partially supported by the National Natural Science Foundation of China (Grant Nos. 61433014, 61803073 and 11975071) and by the Fundamental Research Funds for the Central Universities under Grant Nos. ZYGX2016J196.

\biboptions{numbers,sort&compress}

\end{document}